# A Dual-Tuned Concentric Multimodal RF Coil for 7T ¹H/³¹P MRSI: Concurrently Enhancing B1 Efficiency Over Single-Tuned References

Yunkun Zhao[1], Xiaoliang Zhang[1,2*]

[1]Department of Biomedical Engineering, [2]Department of Electrical Engineering, State University of New York at Buffalo, Buffalo, NY, United States

*Corresponding author:

Xiaoliang Zhang, Ph.D.
Bonner Hall 215E
Department of Biomedical Engineering
State University of New York at Buffalo
Buffalo, NY, 14226
U.S.A.

Email: xzhang89@buffalo.edu

**Abstract**

This study presents the design, simulation, and experimental validation of a dual-tuned concentric multimodal surface coil for 7T 1H/31P magnetic resonance spectroscopic imaging (MRSI), developed to significantly enhance 31P B1 efficiency while improving 1H performance. The coil architecture utilizes two interleaved sets of three concentric loop resonators. Intra-nucleus electromagnetic coupling within each three-loop set generates a spectrum of eigenmodes; the operational modes for 1H and 31P were specifically selected because their co-directed current distributions reinforce the magnetic field at the center, yielding B1 patterns that resemble those of conventional single-loop surface coils but with superior efficiency. Full-wave electromagnetic simulations and bench measurements on a fabricated prototype were conducted to characterize the multimodal resonance behavior, scattering parameters, B1 distribution, and 10-g local SAR, using size-matched conventional single-tuned loops as references. The results confirmed that the design reproducibly generated the predicted eigenmode ordering with sufficient spectral separation to prevent interference from parasitic or undesired modes. Notably, the multimodal design achieved an 83% boost in 31P B1 efficiency and a 21% boost in 1H B1 efficiency at the coil center compared to same-sized single-tuned references. Sufficient inter-nuclear decoupling was achieved to prevent signal leakage between channels, and simulations with a human head model confirmed that the peak 10-g local SAR remained comparable to conventional designs. These findings demonstrate that this multimodal concentric design offers a robust and highly efficient solution for multinuclear MRSI at ultrahigh fields, effectively mitigating the sensitivity limitations of X-nuclei without compromising proton-based imaging capabilities. Furthermore, the proposed design can serve as a scalable building block for the development of high-density, multichannel dual-tuned RF coil arrays, provided appropriate inter-channel decoupling strategies are applied.

# 1. Introduction

Heteronuclear magnetic resonance (MR) serves as a powerful, non-invasive imaging tool to depict the metabolic processes of living systems [1-6]. Among various X-nuclei, 31P is a principal target because it allows for the interrogation of high-energy phosphate compounds, providing a direct window into cellular energetics in vivo [7-10]. However, heteronuclear MR is fundamentally limited by a low sensitivity problem, driven by the low natural abundance and low Larmor frequency of heteronuclei, which results in poor spatial resolution and limited spectral dispersion. While ultrahigh-field (UHF) strengths of 7T and above offer clear advantages in increasing the signal-to-noise ratio (SNR) and spectral separation [11-17], they also impose stricter requirements on the underlying radiofrequency (RF) hardware [18-33].

To successfully perform 31P MRSI at 7T, a dual-tuned 1H/31P RF coil is highly desired and, often, essential . The 31P channel is required for metabolic data acquisition, while the 1H channel is indispensable for anatomical localization and B0 shimming. Despite their necessity, current dual-tuned coil designs often suffer from performance degradation. Traditional methods, such as trap-based single loops[34-38], nested or concentric loop pairs[39-43], microstrip and transmission-line implementations[44-56], and modified birdcage-style structures[57-59], frequently decrease the B1 efficiency of both the heteronuclear and proton channels compared to their single-tuned counterparts. Furthermore, achieving sufficient inter-nuclear decoupling, reaching the high resonant frequencies required for 1H at 7T (~300 MHz), and being structurally compact remains a significant engineering challenge.

A promising solution to these challenges lies in multimodal resonator design based on intentional electromagnetic coupling [60-64]. When multiple conductors are placed in proximity, they form a system with a discrete set of eigenmodes [65, 66]. If the geometry is optimized, a specific eigenmode can be excited where currents flow in-phase (co-directed) across all conductors, reinforcing the magnetic field at the center. However, such systems also produce "parasitic" or undesired eigenmodes where counter-directed currents lead to B1 cancellation. Therefore, a successful multimodal design must ensure that the operational, high-efficiency mode is well-isolated from these parasitic resonances.

In this work, we propose a dual-tuned 7T 1H/31P concentric surface coil designed to improve B1 efficiency for both channels. The architecture consists of two interleaved sets of three concentric loop resonators of varying diameters. By leveraging mutual electromagnetic coupling, each set generates three distinct eigenmodes. We specifically selected the lowest-frequency mode for each nucleus because its co-directed current distribution provides a significant boost in B1 efficiency while maintaining a profile suitable for MR applications.

We evaluate the proposed design using full-wave electromagnetic simulations to characterize S-parameters, $B_1^+$ efficiency, and SAR, complemented by experimental measurements of a fabricated prototype involving standard RF bench testing and field evaluation via a high-resolution 3D B-field mapping system. We demonstrate that the design successfully achieves sufficient inter-nuclear decoupling while significantly outperforming conventional designs. Specifically, we report an ~83% boost in 31P B1 efficiency and a ~21% boost in 1H B1 efficiency at the coil center compared to same-sized single-tuned coils. Furthermore, we verify that the parasitic modes are spectrally far enough from the operational frequencies to ensure stability, and that the SAR profile remains comparable to standard designs. This work provides a practical and highly efficient approach toward multinuclear MRSI at ultrahigh fields. Beyond providing a standalone solution, this design can potentially serve as a scalable building block for high-density, multichannel dual-tuned arrays, addressing the persistent need for high-sensitivity multinuclear imaging over larger volumes, if appropriate inter-channel decoupling strategies are applied [67-78].

## 2. Methods

### 2.1  EM simulation

Figure 1 illustrates the dual-tuned concentric surface coil. The device comprises six coplanar circular loops arranged on a common plane: three $^1$H elements (diameters 10, 9, and 8 cm) and three $^{31}$P elements (9.5, 8.5, and 7.5 cm). All conductors were formed from 16-AWG copper. Each individual $^1$H loop was first tuned near 330 MHz and each $^{31}$P loop near 160 MHz. Within each nucleus, intentional magnetic coupling yields a discrete set of eigenmodes; the operational imaging modes used here occur near 300 MHz ($^1$H) and 120 MHz ($^{31}$P) and exhibit co-directed currents across the concentric elements. The outer $^1$H loop was driven and matched at 300 MHz. Each $^{31}$P loop incorporated an LC trap resonant near 330 MHz to suppress off-resonant $^1$H current while preserving intra-nucleus coupling at $^{31}$P. To contextualize performance, size-matched single-loop surface coils were realized for each nucleus in both simulation and bench experiments. The $^1$H reference was a 10 cm diameter loop; the $^{31}$P reference was a 9.5 cm loop. Each was implemented with the same conductor type and substrate support as the concentric design, tuned to its respective Larmor frequency, and matched to 50 Ω.

Full-wave simulations were performed in CST Studio Suite. For both nuclei, we computed S-parameters, $B_1^+$ efficiency, and 10-g local SAR. Fields were normalized to 1 W of accepted power at the driven port. The simulated environment included a homogeneous phantom and a human head bio-model to assess depth performance and safety under realistic loading. Eigenmodes were identified by sweeping frequency and inspecting current directions and field patterns across the concentric elements; the uniform-field modes (Mode 1 for $^{31}$P, Mode 4 for $^1$H) were selected for subsequent analysis. Reference single-loop coils were simulated under identical boundary conditions and post-processing to enable direct comparison. For all cases, we recorded return loss ($S_{11}$), coupling behavior, $B_1^+$ maps on orthogonal planes (Y–Z, X–Z, and X–Y), and peak/average 10-g SAR.

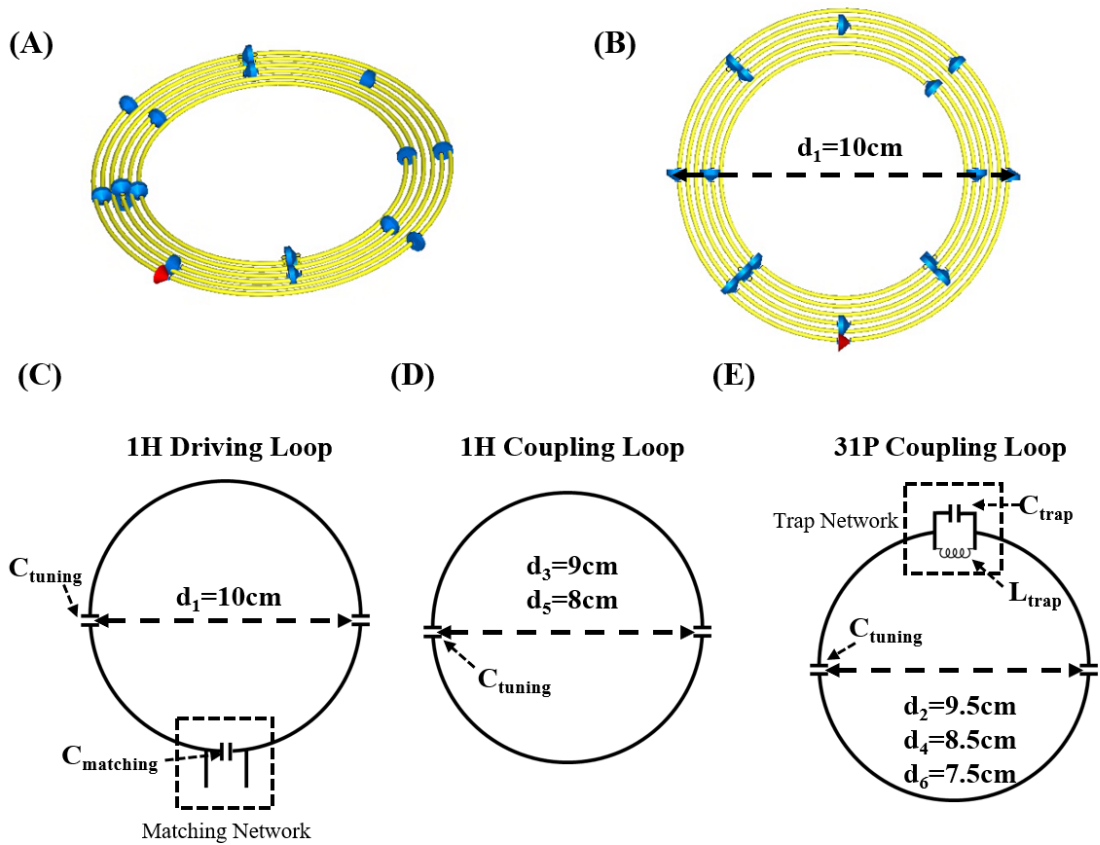

**Figure 1.** Dual-tuned concentric coil for 7 T. (A) Single-loop current concept. (B) Six-loop geometry with outer $^1$H drive ($d_1$=10 cm). (C) $^1$H drive loop (tuning/matching). (D) $^1$H coupling

loops (d₃=9 cm; d₅=8 cm). (E) ³¹P coupling loops (d₂=9.5 cm; d₄=8.5 cm; d₆=7.5 cm) with 330 MHz trap (LC trap). Individual loops tuned around 330 MHz (¹H) and 160 MHz (³¹P); coupling forms operational eigenmodes at around 300 MHz and 120 MHz.

## 2.2 Model Construction and Bench Test

Figure 2A shows the bench-test model of the dual-tuned concentric coil; Figure 2B and 2C show the bench-test models of the size-matched reference surface coils for ³¹P and ¹H. Physical prototypes mirrored the simulated dimensions and component placements. Fixed and variable capacitors were used for fine tuning and matching at each nucleus. Trap networks on the ³¹P elements were implemented as series L–C resonators adjusted to the ¹H frequency. Components were mounted on a non-conductive 3D-printed frame to maintain element spacing and alignment. The driven ¹H port included a compact matching network enclosed for mechanical stability. Bench characterization used a vector network analyzer (VNA) and a 3-D B-field mapping system with a calibrated sniffer probe. For each nucleus, the coil under test was tuned and matched at its operational eigenmode. Accepted power was monitored and used to normalize all maps to 1 W. Spatial scans were acquired over a fixed field-of-view centered on the coil in three orthogonal planes (Y–Z, X–Z, and X–Y). The same procedure and scan region were applied to the single-loop references. Simulation and bench maps were exported and post-processed in a common workflow. B₁ efficiency (µT/ √ W) was computed from the B1 field with accepted-power normalization. The dual-tuned concentric coil is compared against the size-matched single-loop references; comparisons are presented nucleus-wise (¹H vs ¹H reference; ³¹P vs ³¹P reference).

(A) 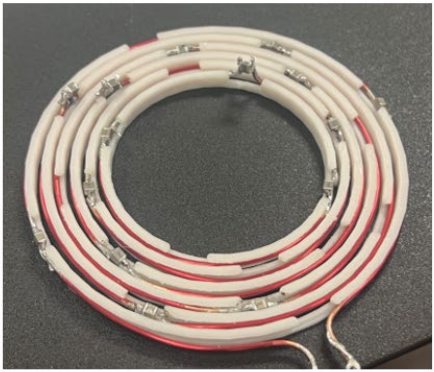

(B) 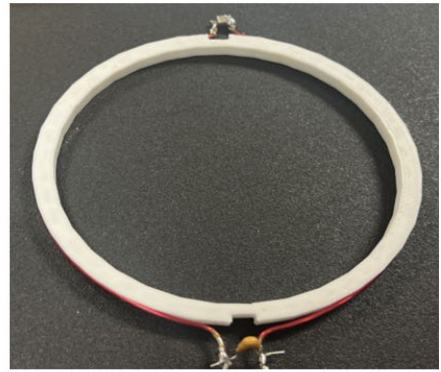

**Figure 2.** (A) Dual-tuned concentric coil (six loops: $^1$H 10/9/8 cm; $^{31}$P 9.5/8.5/7.5 cm). (B) $^{31}$P reference single loop (9.5 cm). (C) $^1$H reference single loop (10 cm). All built from 16-AWG copper on the same non-conductive frame.

## 3. Results

### 3.1 Simulated Resonant Frequency and Field Distribution

In Figure 3, the simulated $S_{11}$ spectrum exhibits two sets of three well-separated resonances that correspond to the coupled eigenmodes of the dual-tuned concentric coil. Modes 1–3 appear in the lower-frequency cluster associated with the $^{31}$P channel, and Modes 4–6 lie in the higher-frequency cluster associated with the $^1$H channel. The red dashed boxes in Figure 3 mark Mode 1 ($^{31}$P) and Mode 4 ($^1$H), which were identified a priori as imaging candidates because they were expected to support coherent current flow across the concentric elements. The depth and isolation of the $S_{11}$ minima indicate stable matching and sufficient spectral separation, enabling selective operation of the desired mode without multi-port feeds or additional decoupling networks. In addition to the six imaging-relevant modes, a cluster of high-frequency 'trap modes' appears near ~500 MHz, arising from the $^{31}$P-loop LC traps tuned to the $^1$H band; these fall outside our bands of interest and were excluded from analysis [79].

In Figure 4A, the simulated vector B-field direction maps are shown for all six resonances. For Mode 1 and Mode 4, the arrows across the three concentric loops are co-directed, producing a uniform, constructive $B_1$ pattern that reinforces the field in the central region and is therefore suitable for imaging. By contrast, Modes 2–3 ($^{31}$P) and Modes 5–6 ($^1$H) display segments with opposing directions and nodal regions between elements, indicating partial cancellation and reduced central $B_1^+$ relative to the uniform modes. In Figure 4B, the simulated B1 vector fields are plotted for all six resonances on the same Y–Z plane used in Figure 4A. Mode 1 ($^{31}$P) and Mode 4 ($^1$H) show co-directed vectors across the three concentric loops, indicating a uniform current phase that supports the single-lobe magnitude patterns observed in Figure 4A. By contrast, Modes 2–3 ($^{31}$P) and Modes 5–6 ($^1$H) display counter-rotating segments and nodal lines between elements; these phase reversals explain the split-lobe/low-magnitude patterns in Figure 4A and make these modes unsuitable for signal excitation and reception. These vector maps therefore justify selecting Mode 1 and Mode 4 as the operational imaging modes used in subsequent efficiency comparisons.

In Figure 5, the dual-tuned concentric coil operating in the imaging modes shows strong, centrally reinforced B1 fields for both nuclei. For $^{31}$P (left), $B1^+/B1^-$ maps exhibit a compact central maximum with smooth depth fall-off, consistent with a well-formed surface-coil pattern. For $^1$H (right), the maps show a similar compact central maximum and smooth decay, which comparable to a normal single-loop surface coil and confirming that the selected eigenmode reproduces the expected surface coil behavior without introducing peripheral hot spots. Across planes, $B1^+$ and $B1^-$ closely agree for both nuclei.

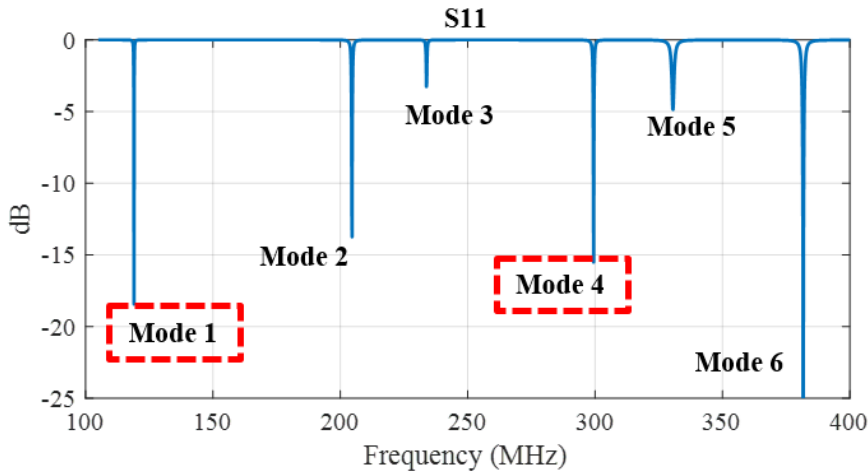

**Figure 3.** Simulated $S_{11}$ of the dual-tuned concentric coil showing six resonances: Modes 1–3 ($^{31}$P) at lower frequencies, Modes 4–6 ($^1$H) higher. Red dashed boxes mark the imaging modes (Mode 1, $^{31}$P; Mode 4, $^1$H).

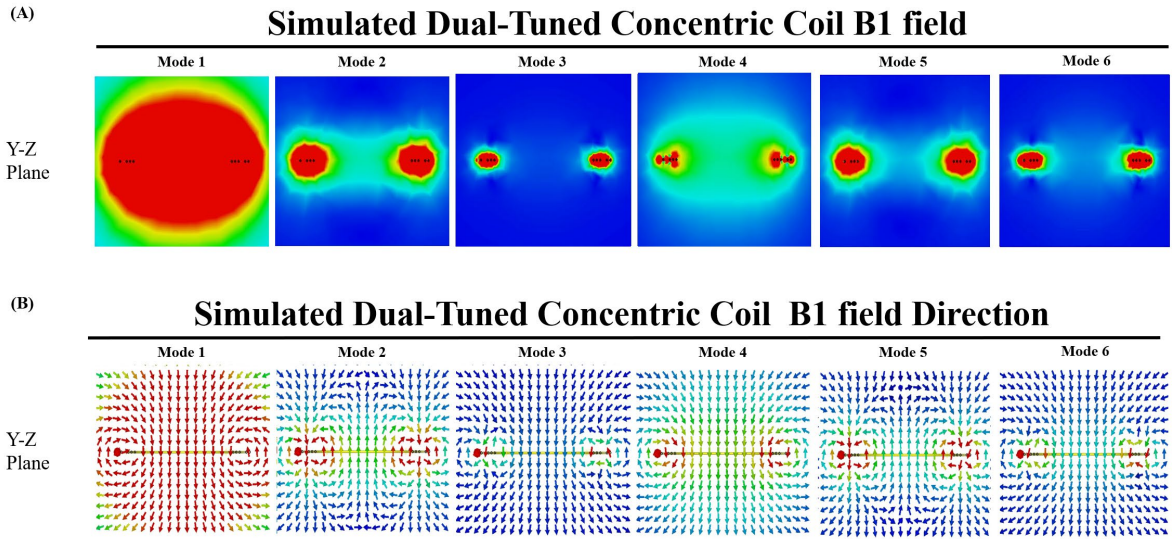

**Figure 4.** (A) Simulated B1 magnitude (Y–Z plane) for Modes 1–6. Modes 1 ($^{31}$P) and 4 ($^1$H) form single-lobe patterns; other modes show split/weak lobes. (B) Simulated B1 vector direction (Y–Z plane) for Modes 1–6. Modes 1 and 4 have co-directed vectors; other modes show phase

reversals.

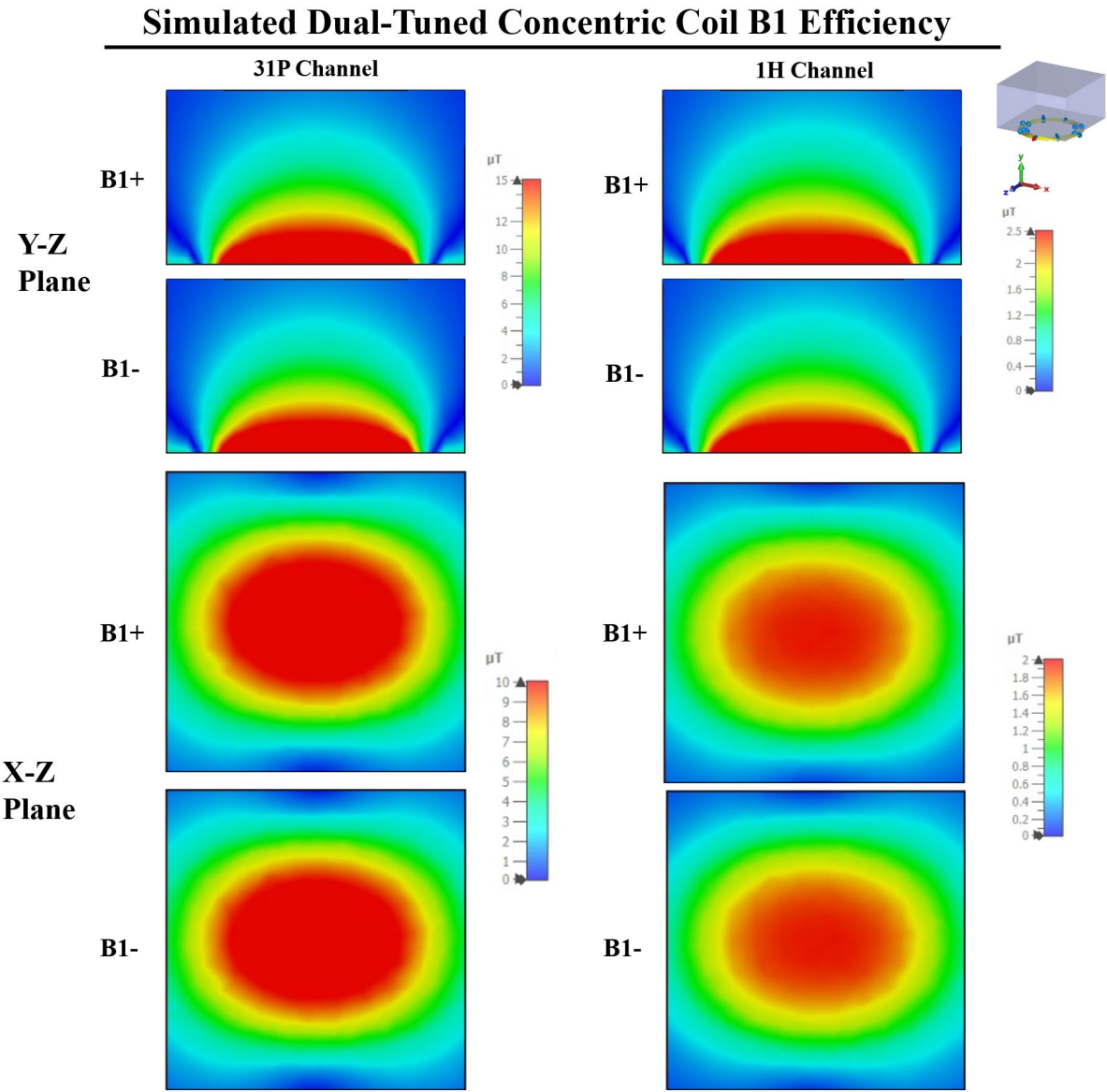

**Figure 5.** Simulated B1 efficiency (1 W accepted power) for imaging modes: $^{31}$P (left) and $^{1}$H (right).

## 3.2 Measured Scattering Parameters and Field Distribution

In Figure 6, the benchtest $S_{11}$ spectrum exhibits six distinct resonances with the same ordering and separation as in simulation: three $^{31}$P modes at lower frequencies followed by three $^{1}$H modes at higher frequencies. The two operational imaging modes, Mode 1 and Mode 4 are adequately matched. The close agreement in frequency ordering and inter-mode spacing with the simulated response indicates that the eigenmode placement is robust to hardware non-idealities. Minor shifts in absolute frequency are consistent with component tolerances, solder parasitics, cable loading, and the trap networks on the $^{31}$P elements. Overall, the bench data confirm that single-port excitation can reliably select the uniform-direction modes identified in simulation.

In Figure 7, bench B1 efficiency maps across the X-Y, X-Z, and Y-Z planes are shown for the $^{31}$P imaging mode (left) and the $^{1}$H imaging mode (right), normalized to 1 W accepted power. B$_1$ measurements were acquired 1 cm above the coil in the X–Y and Y–Z planes, and 3 cm above the coil in the X–Z plane, to assess near-field performance. In both Y–Z and X–Z planes, the patterns exhibit a compact peak at the coil vicinity with smooth depth fall-off, which is the expected surface coil like profile. A slight asymmetry in the $^{1}$H maps is attributable to cable proximity but does not affect the overall surface coil like behavior. These measurements agree with the simulations.

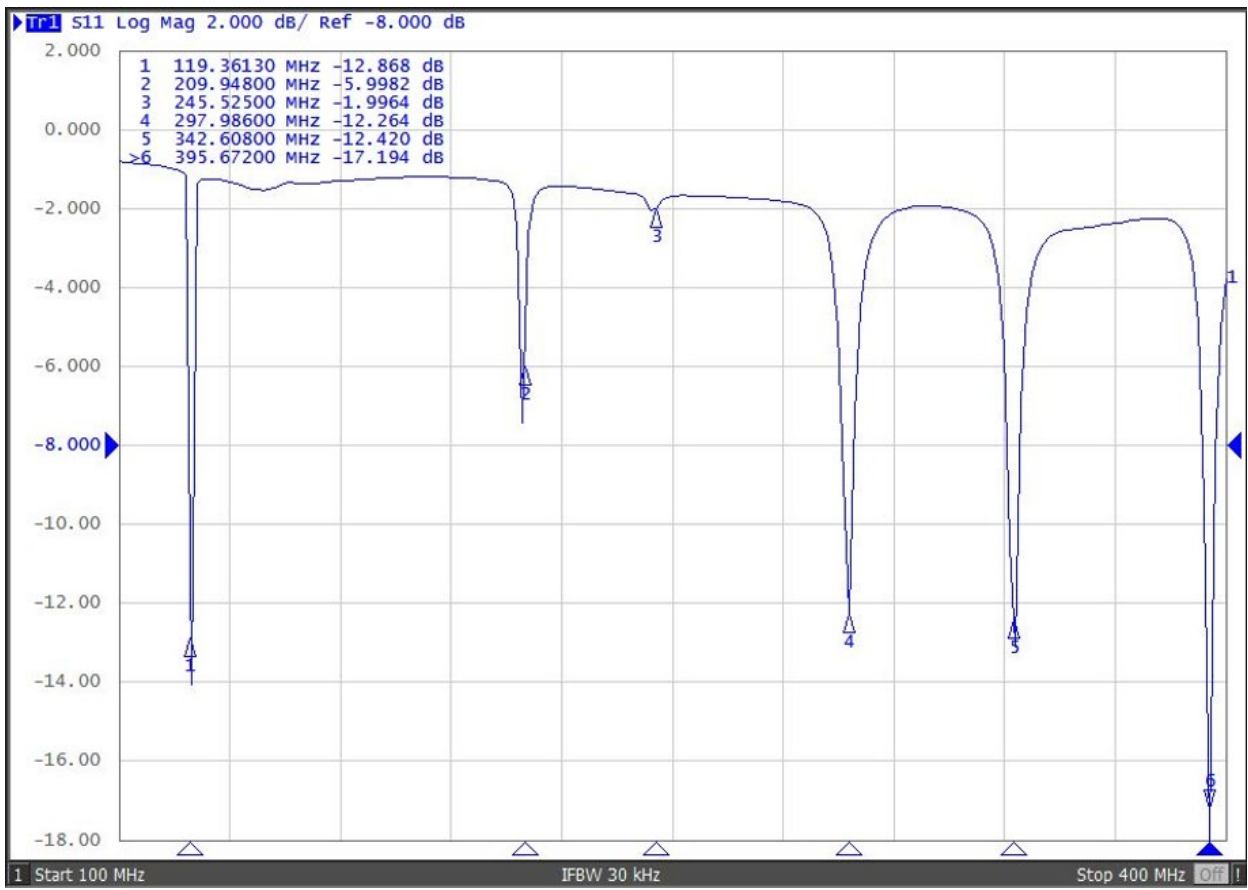

**Figure 6.** Measured S$_{11}$ from 100–400 MHz with six resonances matching simulation.

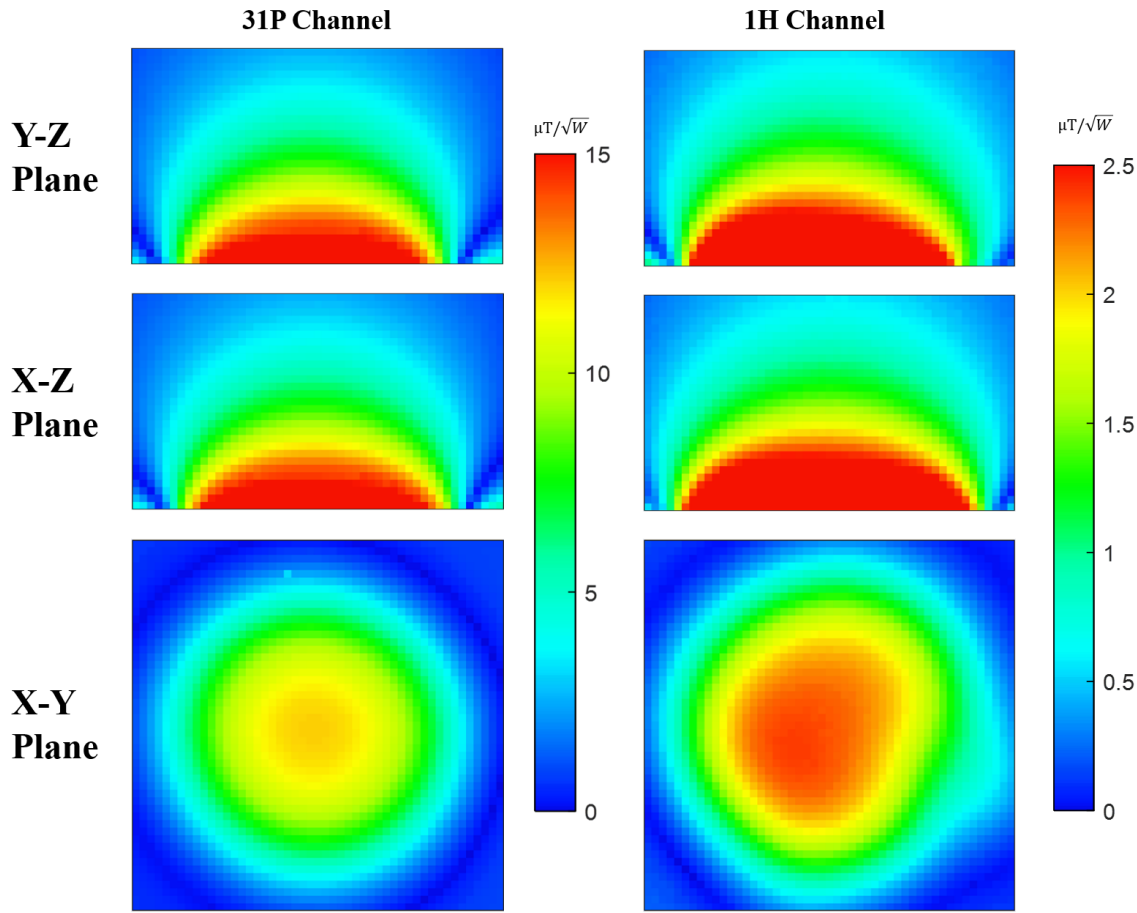

**Figure 7.** Measured B1 efficiency of the dual-tuned concentric coil (normalized to 1 W accepted power). Left: $^{31}$P imaging mode. Right: $^{1}$H imaging mode.

## 3.3 Field Distribution and Efficiency Evaluation

In Figure 8A, simulated B1 maps (normalized to 1 W accepted power) compare the dual-tuned concentric coil with a conventional single-loop for both nuclei. For $^{31}$P, the concentric design yields a visibly stronger central $B_1^+/B_1^-$ than the single-tuned reference loop. For $^{1}$H, the concentric coil shows a surface coil like pattern that is comparable to the single-tuned loop in both magnitude and lateral spread. In Figure 8B, depth and lateral line profiles confirm these observations: the concentric coil delivers substantially higher $^{31}$P B1 efficiency across relevant depths, while $^{1}$H B1 efficiency remains comparable to the conventional surface coil. At the coil center, we measured an ~83% boost in 31P B1 efficiency and a ~21% boost in 1H B1 efficiency compared to same-sized single-tuned coils. Together, these data demonstrate that the proposed geometry improves $^{31}$P B1 efficiency without compromising $^{1}$H performance.

In Figure 9A, simulated B1 efficiency is compared for $^{31}$P and $^{1}$H in a human head bio-model. For $^{31}$P, the dual-tuned concentric coil shows higher B1 efficiency than the single loop in both Y–Z and X–Z planes, with a clearer central maximum at depth. For $^{1}$H, the concentric coil provides comparable B1 efficiency to the reference loop near the target region. In Figure 9B, 10-g local SAR for the $^{1}$H channel exhibits similar peak values and locations for both designs, indicating comparable safety behavior under matched drive.

In Figure 10, measured B1 field efficiency maps compare the dual-tuned concentric coil with a conventional single loop for both nuclei. For $^{31}$P (left panel), the concentric coil shows a higher central B1 in Y–Z, X–Z, and X–Y planes than the reference. For $^{1}$H (right panel), the concentric coil provides comparable B1 to the single loop with a similar surface-coil profile. These measurements agree with the simulations, confirming improved $^{31}$P B1 efficiency while preserving $^{1}$H performance.

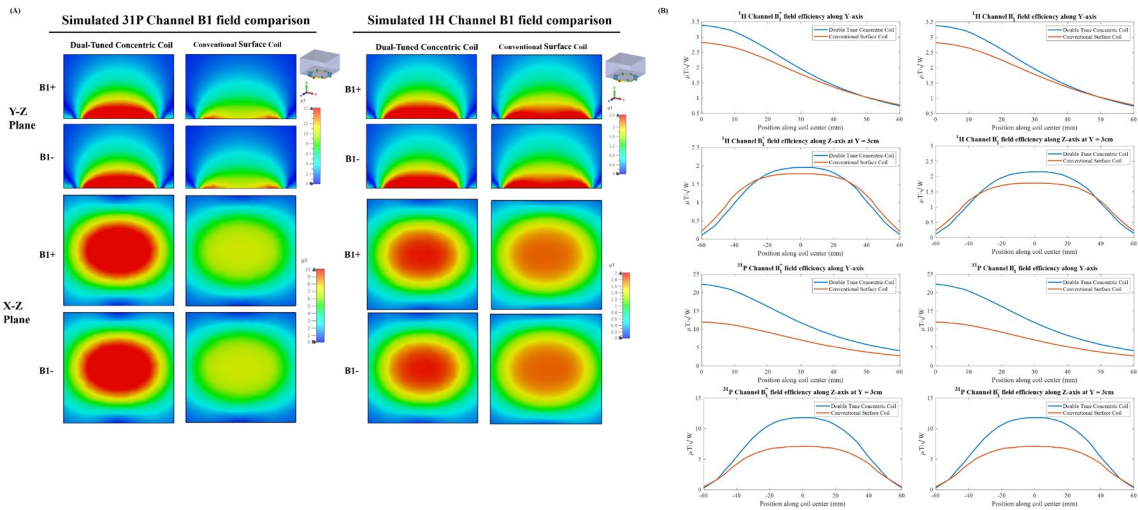

**Figure 8.** Simulated B1 efficiency comparison with a conventional single loop (1 W accepted power). (A) $^{31}$P (left) and $^{1}$H (right) maps: the concentric coil provides higher $^{31}$P B1 efficiency and comparable $^{1}$H efficiency. (B) Line profiles along depth and lateral directions confirm higher $^{31}$P and comparable $^{1}$H performance.

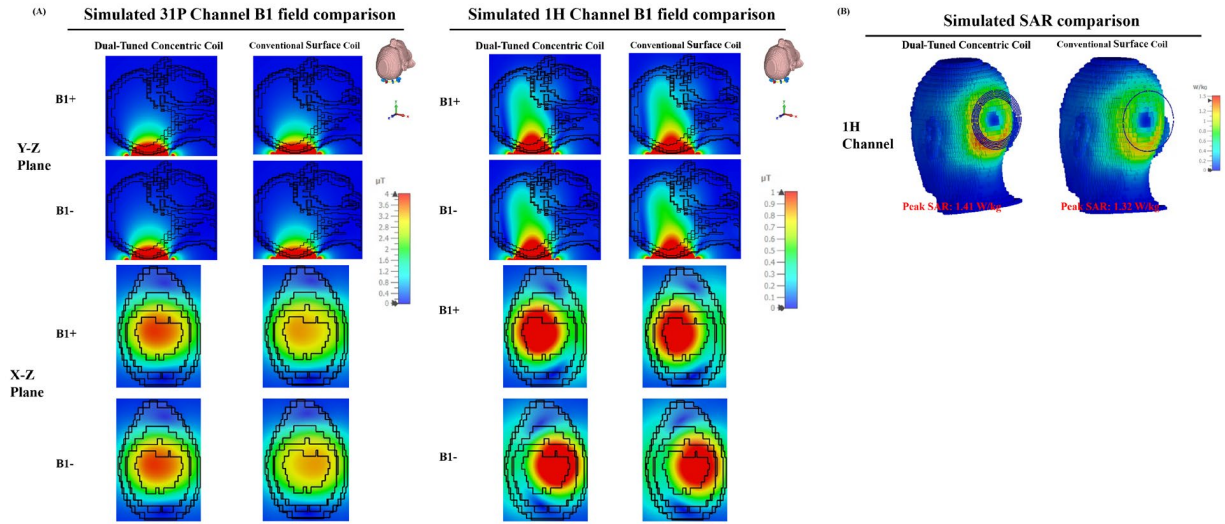

**Figure 9.** Human head bio-model. (A) Simulated B1 efficiency: the concentric coil yields higher $^{31}$P and comparable $^{1}$H efficiency versus a single loop. (B) 10-g local SAR for the $^{1}$H channel.

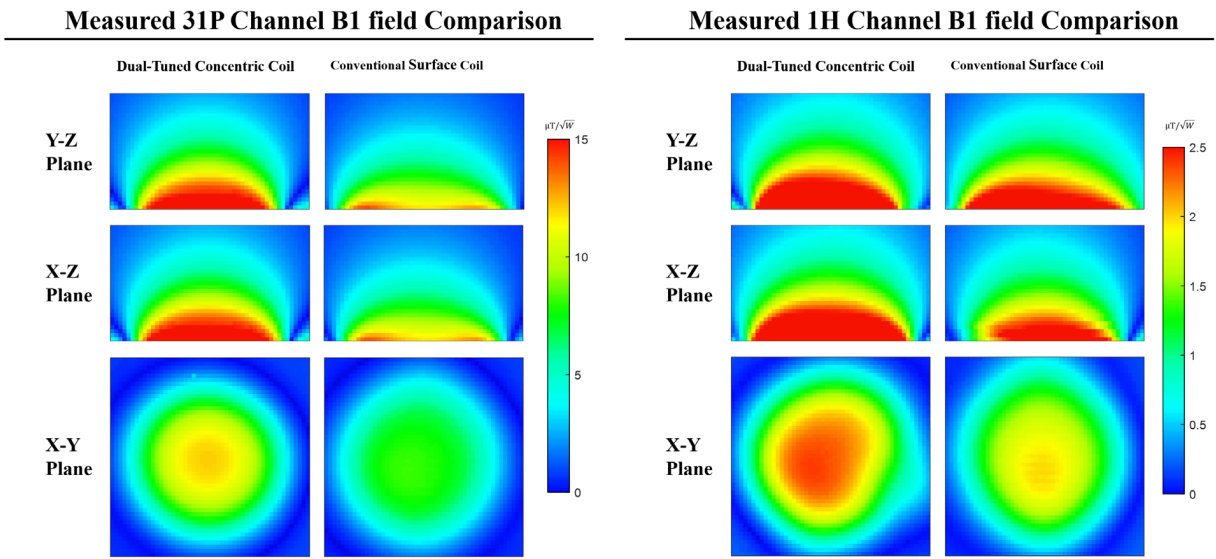

**Figure 10.** Bench B1 efficiency comparison (μT/√W). Measured Y-Z, X-Z, and X-Y planes for $^{31}$P (left) and $^{1}$H (right).

## 4. Conclusion

In this work, we introduced a dual-tuned multimodal resonator technique and developed and validated a dual-tuned 7T 1H/31P surface coil based on the concentric multimodal resonator architecture. By leveraging the principle of intentional electromagnetic coupling, we demonstrated that a specific eigenmode, characterized by co-directed currents across three concentric loops, can be successfully tuned and matched for high-field MR applications. This multimodal approach effectively mitigates the fundamental sensitivity limitations of heteronuclear imaging. Specifically, our results confirm that this design provides an ~83% boost in 31P B1 efficiency and a ~21% boost in 1H B1 efficiency at the coil center compared to conventional single-tuned loops of the same size.

Beyond efficiency gains, high inter-nuclear isolation was achieved between the 1H and 31P channels, preventing the signal leakage and sensitivity loss typically encountered in conventional trap-based or nested designs. Furthermore, the management of mutual electromagnetic coupling among the concentric loops ensured that the operational high-efficiency modes remained spectrally isolated from parasitic resonances, providing robust tuning stability. Safety evaluations via numerical simulation confirmed that the multimodal current distribution maintains a comparable SAR profile to standard single-loop resonators.

Ultimately, this concentric multimodal dual-tuned design offers a potent combination of improved sensitivity for both 31P and 1H and reliable anatomical referencing for multinuclear MRSI at ultrahigh fields. Because of its compact and efficient nature, this design can serve as a scalable building block for the development of high-density, multichannel dual-tuned RF coil arrays, provided appropriate inter-channel decoupling strategies are applied. Addressing the persistent need for high-sensitivity and high-speed multinuclear imaging over larger volumes remains a key objective for future UHF research.

## 5. Discussion

The results presented in this study demonstrate that a concentric multimodal resonator design can effectively mitigate the sensitivity trade-offs traditionally associated with dual-tuned RF hardware. The most significant finding is the substantial enhancement of B1 efficiency for both 31P and 1H relative to size-matched single-tuned conventional loops. This represents a difference from the conventional understanding of multinuclear coil design, where the integration of a second tuning frequency—even when using optimized LC traps—often introduces parasitic resistance or alters current paths in a way that leads to a net efficiency penalty. In our design, while LC traps were utilized in the 31P loops to suppress induced 1H currents, the multimodal architecture compensated for potential losses, resulting in a significant performance gain.

The mechanism behind this performance boost is the constructive interference of co-directed currents across the interleaved concentric loops. By intentionally leveraging mutual electromagnetic coupling, we create a system where the operational eigenmode acts as a distributed resonator. In this mode, the current flows in the same direction across all three loops for the target nucleus, effectively reinforcing the magnetic field not only at the center of the coil but also extending through the imaging depth. This reinforcement effect provides the substantial gain in 31P B1 efficiency, which is particularly critical for addressing the low sensitivity inherent in phosphorus MRSI at 7T.

Our findings confirm that the interleaved geometry is still able to provide sufficient coupling among intra-nuclear loops to separate these modes, ensuring that the operational mode remains stable and unaffected by B1 cancellation or energy leakage into parasitic resonances. Furthermore, the proposed dual-tuned multimodal approach allows for independent excitation and reception with high efficiency. In many dual-tuned designs, "linking" or residual coupling between the 1H and X-nucleus channels leads to signal leakage and wasted power; however, our prototype demonstrates that high inter-nuclear isolation can be achieved simultaneously with enhanced B1 sensitivity.

A major advantage of this concentric multimodal resonator is its compactness. Unlike many existing dual-tuned RF coil designs that rely on bulky 3D structures, this design maintains a 2D, flat profile. This compact footprint makes the proposed design uniquely feasible as a scalable building block for the development of high-density, massive dual-tuned RF coil arrays. Such arrays are essential for enabling fast and highly sensitive heteronuclear MRSI over larger volumes, providing a pathway to translate these sensitivity gains from single-element resonator into high performance multichannel dual-tuned RF arrays.

Finally, the localized SAR profile of the proposed coil remains comparable to standard single-tuned loop resonators. This suggests that the multimodal reinforcement of the B1 field does not come at the cost of increased electric field (E-field) deposition in the tissue. By combining improved sensitivity for both 31P and 1H with a safety profile and physical structure suitable for array integration, this design addresses the persistent need for high-performance multinuclear imaging RF coils at ultrahigh fields.

# Acknowledgments

This work is supported in part by the NIH under a BRP grant U01 EB023829 and by the State University of New York (SUNY) under SUNY Empire Innovation Professorship Award.